\documentclass[usenatbib]{mn2e}
\usepackage{times}
\usepackage{epsfig}

\newcommand{\be}{\begin{equation}}
\newcommand{\ee}{\end{equation}}
\newcommand{\etal}{et al.\ }

\title[Fossil groups and formation of BCGs]
{The central elliptical galaxy in fossil groups and formation 
of BCGs}
\author[Khosroshahi, Ponman \& Jones]{
Habib G. Khosroshahi\thanks{E-mail:
habib@star.sr.bham.ac.uk (HGK)}, Trevor. J. Ponman \& Laurence R. Jones \\
School of Physics and Astronomy, The University of Birmingham,
Birmingham B15 2TT, UK}

\begin{document}

\date{Accepted, Received}

\pagerange{\pageref{firstpage}--\pageref{lastpage}} \pubyear{2004}

\maketitle

\label{firstpage}

\begin{abstract} 

We study the dominant central giant elliptical galaxies in ``Fossil
groups'' using deep optical (R-band) and near infrared (Ks-band)
photometry. These galaxies are as luminous as the brightest cluster
galaxies (BCGs), raising immediate interest in their link to the
formation of BCGs and galaxy clusters. However, despite apparent
similarities, the dominant fossil galaxies show non-boxy isophotes, in
contrast to the most luminous BCGs. This study suggests that the
structure of the brightest group galaxies produced in fossil groups
are systematically different to the majority of BCGs. If the fossils
do indeed form from the merger of major galaxies including late-types
within a group, then their disky nature is consistent with the results
of recent numerical simulations of semi-analytical models which
suggest that gas rich mergers result in disky isophote ellipticals.

We show that fossils form a homogeneous population in which the
velocity dispersion of the fossil group is tightly correlated with the
luminosity of the dominant elliptical galaxy. This supports the
scenario in which the giant elliptical galaxies in fossils can grow to
the size and luminosity of BCGs in a group environment. However, the
boxy structure of luminous BCGs indicate that they are either not
formed as fossils, or have undergone later gas-free mergers within the
cluster environment.

\end{abstract}


\section{Introduction}

It is believed that most of the large and luminous elliptical galaxies
have formed via mergers of disk galaxies
\citep{toomre72,searle73}. This has been suggested by morphology density
relation \citep{dressler80}, the observed frequency of merging
galaxies at high redshift and also extensively in computer simulations
\citep{barnes89}. Many of these luminous
ellipticals ($M_B\le$ -21) are found in rich galaxy clusters. This,
however, does not imply that they are formed in cluster
environment. The brightest cluster galaxies (BCGs) are of special
interest as they reside close to the centroid of cluster X-ray
emission \citep{jf84}, and the centre of the dark matter distribution
in clusters, as inferred from gravitational lensing \citep{smith05} --
implying that they lie at the minimum of the cluster potential
well. They also show various correlations with cluster properties.

In general, two main formation modes could be assumed for the
hierarchical formation of BCGs according to their formation
environment: {\bf 1)} BCGs formed in the high velocity environment of
clusters. Although the effectiveness of dynamical friction in bringing
individual galaxies to the cluster centre via orbital decay will be
reduced by the high velocity dispersion, infalling groups will still
suffer rapid orbital decay if they survive long enough, and can then
deposit their brightest galaxies in the cluster core, where they can
merge and form a bright elliptical galaxy
\citep{mohr04,hausman78}. {\bf 2)} BCGs are formed in the low velocity
environment of groups, where dynamical friction causes the orbits of
individual galaxies to decay, resulting in the merger of all large
galaxies, if the group forms early and is left undisturbed for a
sufficiently long period \citep{dubinski98,ponman94,jones03}. The group 
containing this `ready made' BCG then provides the nucleus around which 
a cluster forms.

Elliptical galaxies show fine structures and are more complex than
originally thought. These structures take the form of hidden
disks, shells and bars, departures from pure elliptical isophotes
\citep{bender88} and variations in radial surface brightness profiles
\citep{kwkm00}, some of which are found to be environment
dependent \citep{habib04}. BCGs formed in the above two modes should
display several observational signatures of how they formed.
A useful probe is provided by galaxy morphology --
isophotal shapes, radial surface brightness profiles and the presence
or not of multiple-nuclei. Different star forming histories are also 
expected. Some of these studies require space
resolution data, but some can be studied using ground based
observations, including the isophotes of elliptical galaxies.

Based on their isophotal shapes, elliptical galaxies can be classified as
disky or boxy \citep{bender88}. Low mass ellipticals, which are fast
rotating, are usually disky isophote galaxies with positive
fourth-order Fourier coefficient $B_4>0$. Some disky-isophote
ellipticals might contain faint disks similar to S0 galaxies
\citep{scorza95}. 
With negative $B_4$, boxy isophote ellipticals are
less rotationally supported. They generally contain flat cores
\citep{faber97,laine03} and show complex internal kinematics
\citep{emsellem04}. The observations of \citet{rest01} show that it 
is very unlikely to find disky ellipticals which are also core galaxies. 
The majority of the BCGs are found to be core galaxies \citep{laine03}.
The distinct observed properties of disky and
boxy isophotes of elliptical galaxies, such as their radio properties
show that they are more than just an artifact of viewing angle or the
projection on the plane of sky \citep{bender89}. 

It is important to understand the origin of isophotal shapes before
they can be used to trace the formation history of ellipticals.
\citet{naab03} performed a large survey of dissipationless 
merger simulations of disk galaxies and found that unequal-mass 
3:1 to 4:1 mergers lead to fast rotating disky ellipticals while 
equal-mass 1:1 to 2:1 mergers produce slowly rotating, pressure-supported 
ellipticals.
However, \citet{khochfar05} showed that the above scenario is not
able to reproduce the observation that the fraction of boxy and disky
ellipticals depends on galaxy luminosity. They argued that equal-mass
mergers lead to boxy ellipticals and unequal-mass mergers produce
disky ellipticals. However, major mergers between bulge-dominated 
galaxies result in boxy ellipticals, independent of the mass ratio, 
while merger remnants that subsequently accrete gas, leading to a
secondary stellar disk with more than 20 per cent of the total stellar
fraction, are always disky. More recently they showed that mergers of 
spiral galaxies alone cannot reproduce the kinematic and photometric 
properties of very massive elliptical galaxies \citep{naab06}, nor can 
they reproduce the observed correlation between isophotal shapes and the 
luminosity of ellipticals. 

Here we study the isophotes of brightest group galaxies (BGGs) 
in fossils.  A brief introduction to fossils is given in section 2 
where we also describe the sample and observations. Our analysis and 
the results are presented in section 3. A discussion and concluding 
remarks are in section 4. We assume $H_0=70$ km~s$^{-1}$~Mpc$^{-1}$ and
$\Omega_m=0.3$ with cosmological constant $\Omega_\Lambda=0.7$
throughout.

\begin{figure}
\center
\epsfig{file=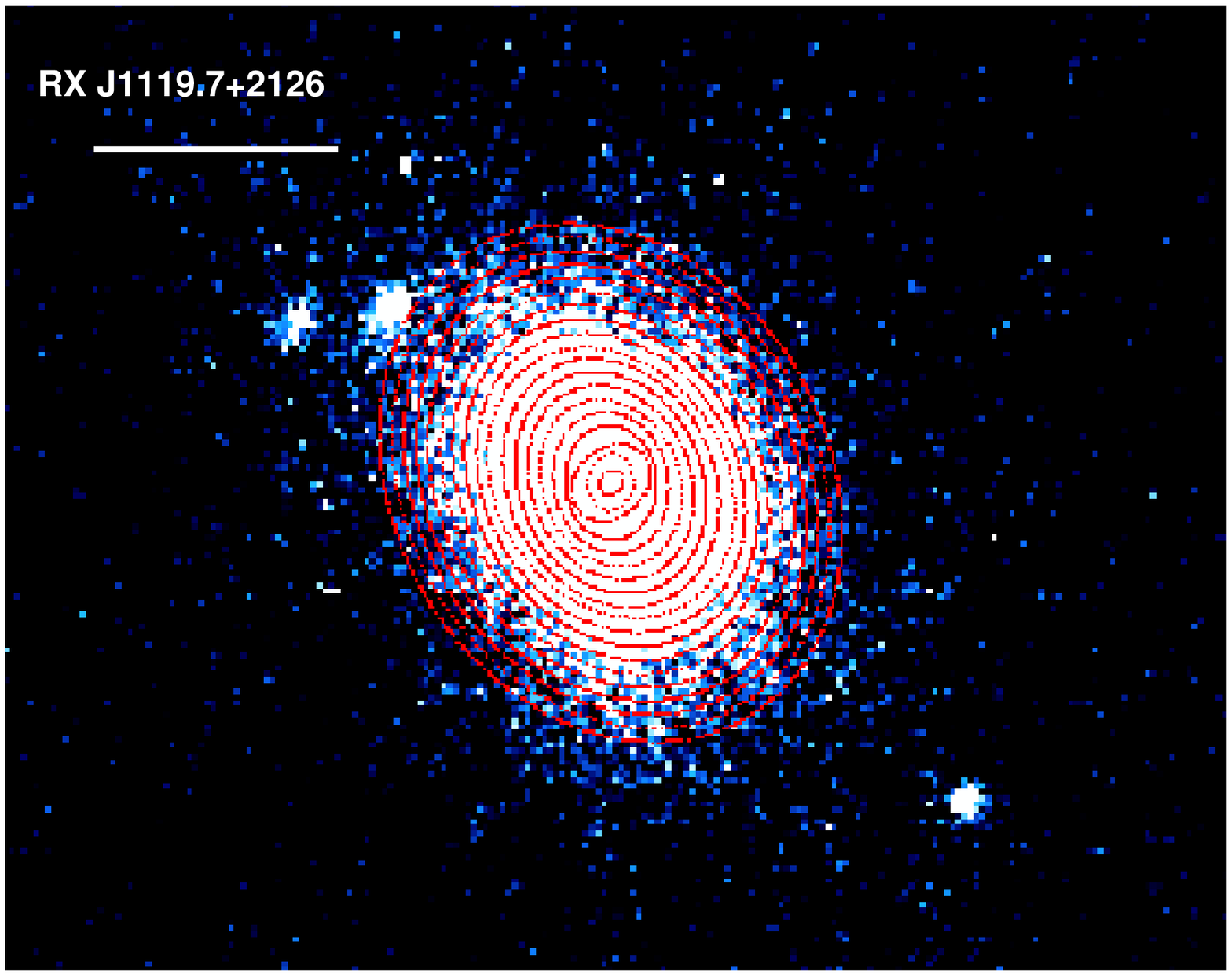,width=1.3in,height=1.2in}
\epsfig{file=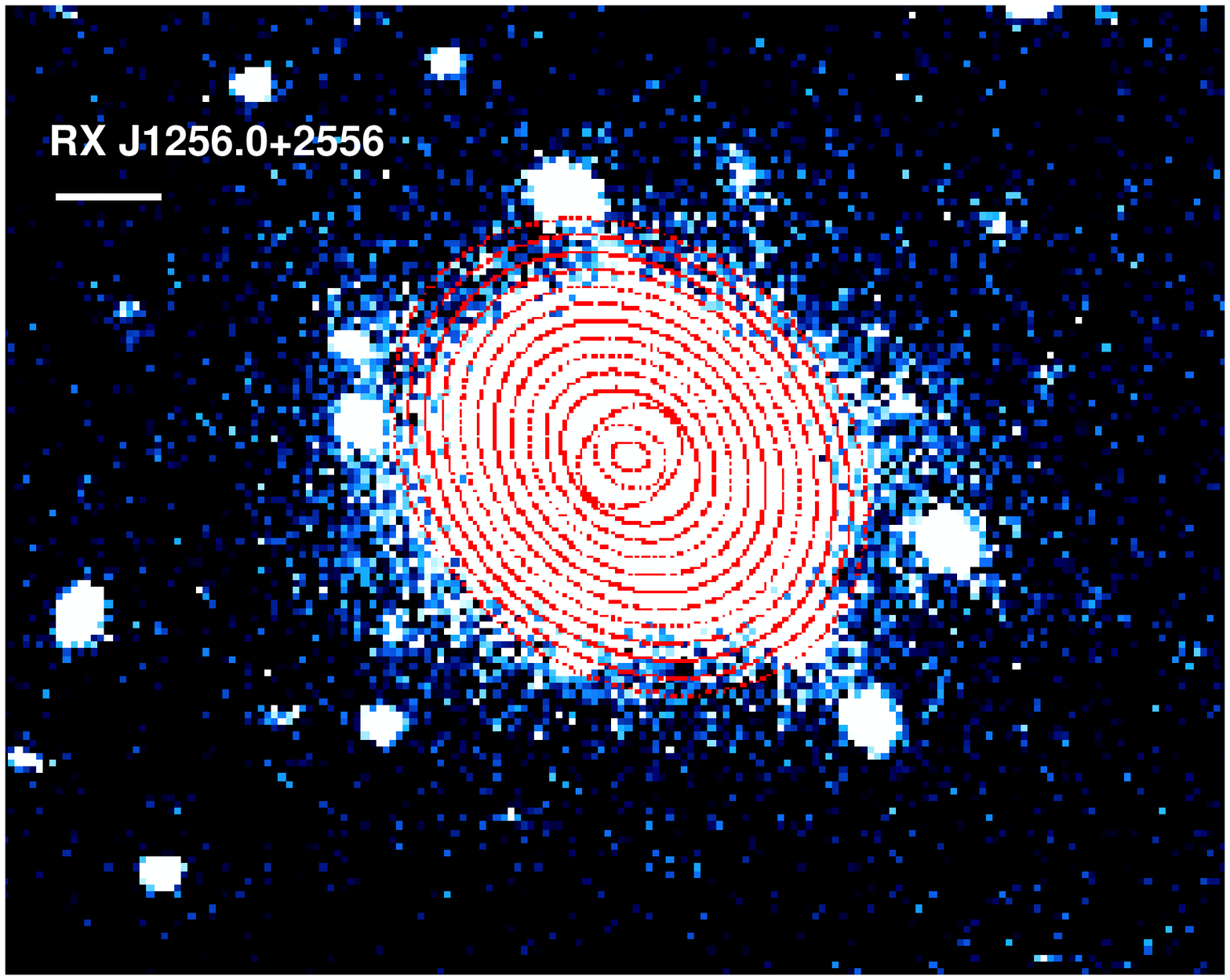,width=1.3in,height=1.2in}
\epsfig{file=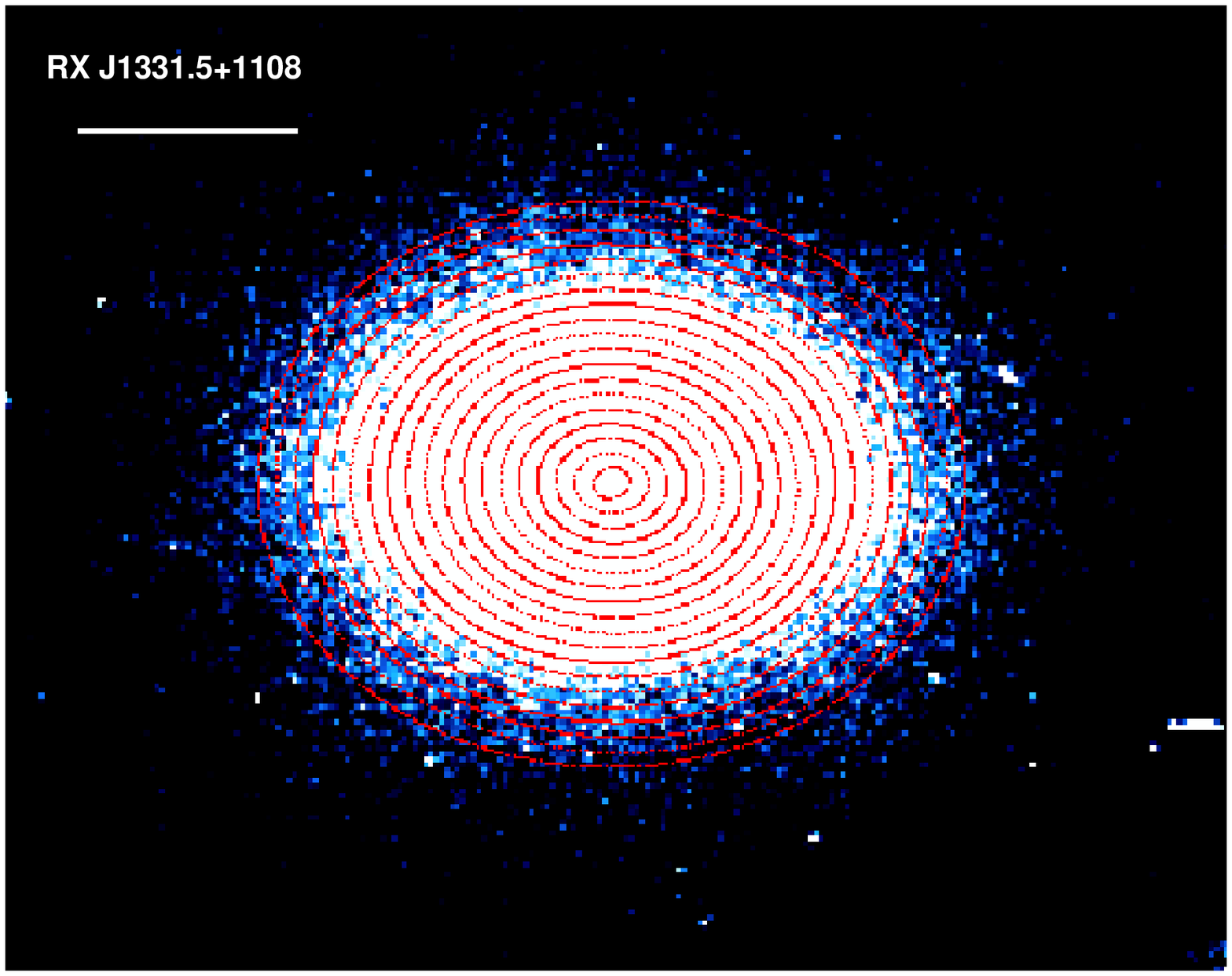,width=1.3in,height=1.2in}
\epsfig{file=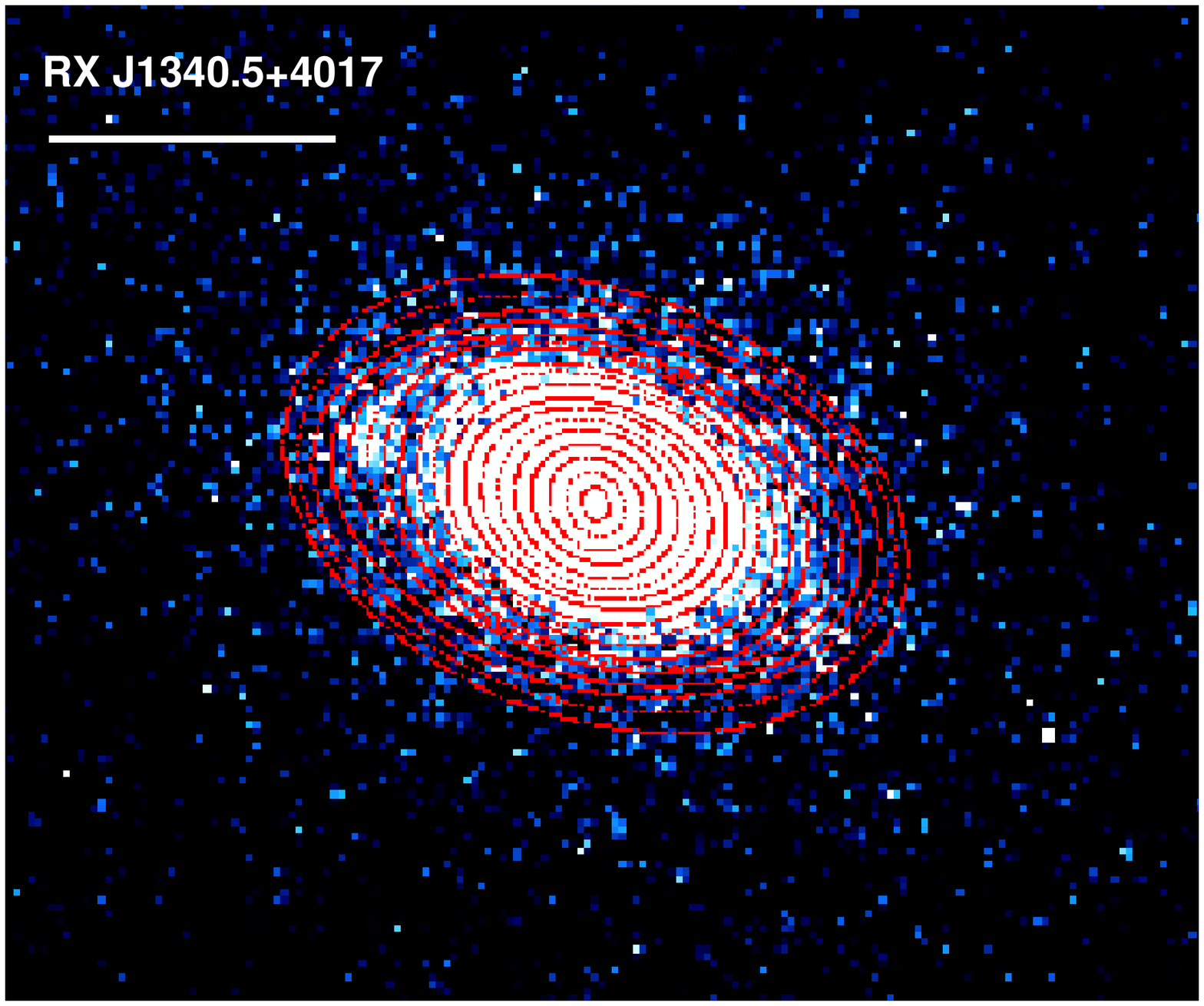,width=1.3in,height=1.2in}
\epsfig{file=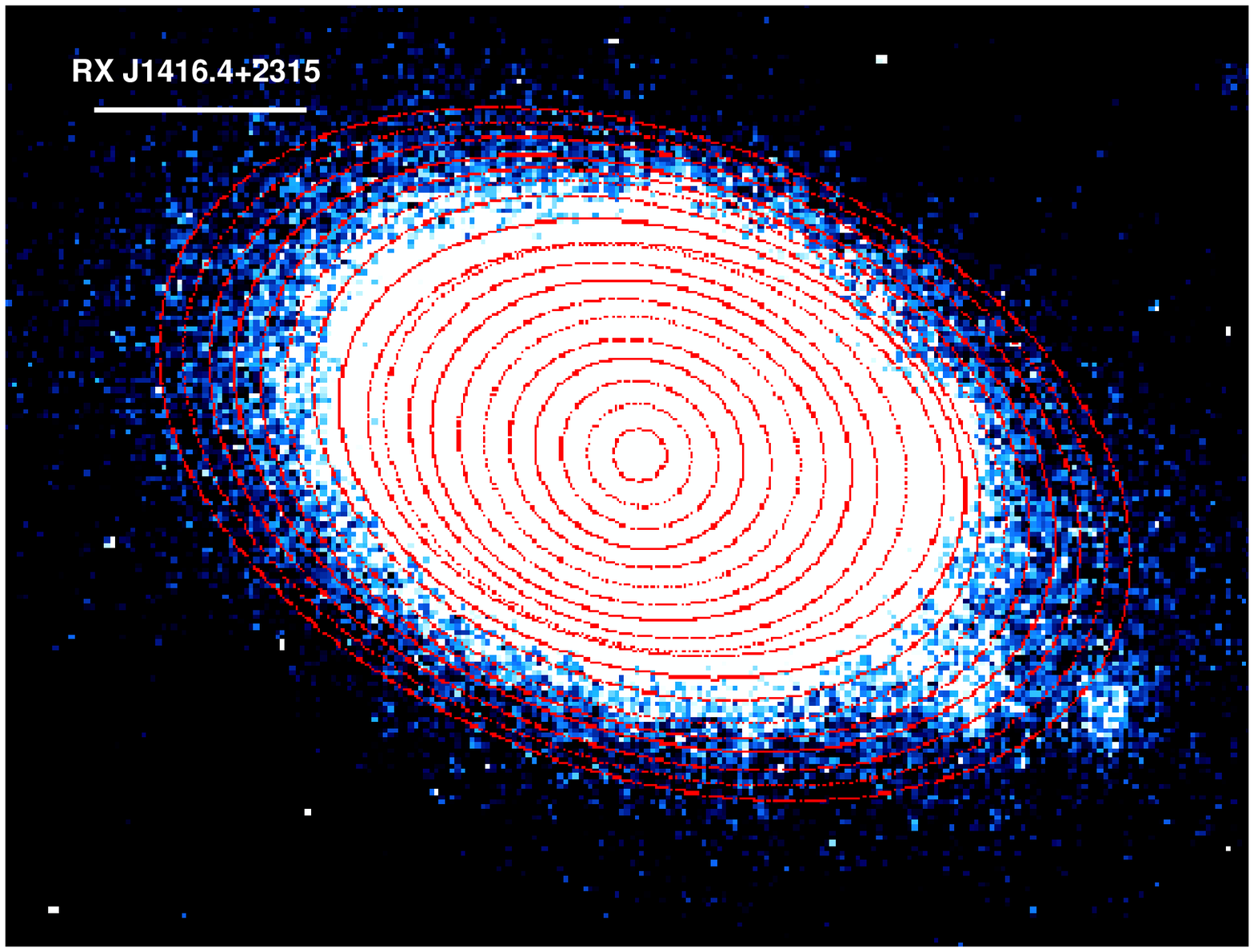,width=1.3in,height=1.2in}
\epsfig{file=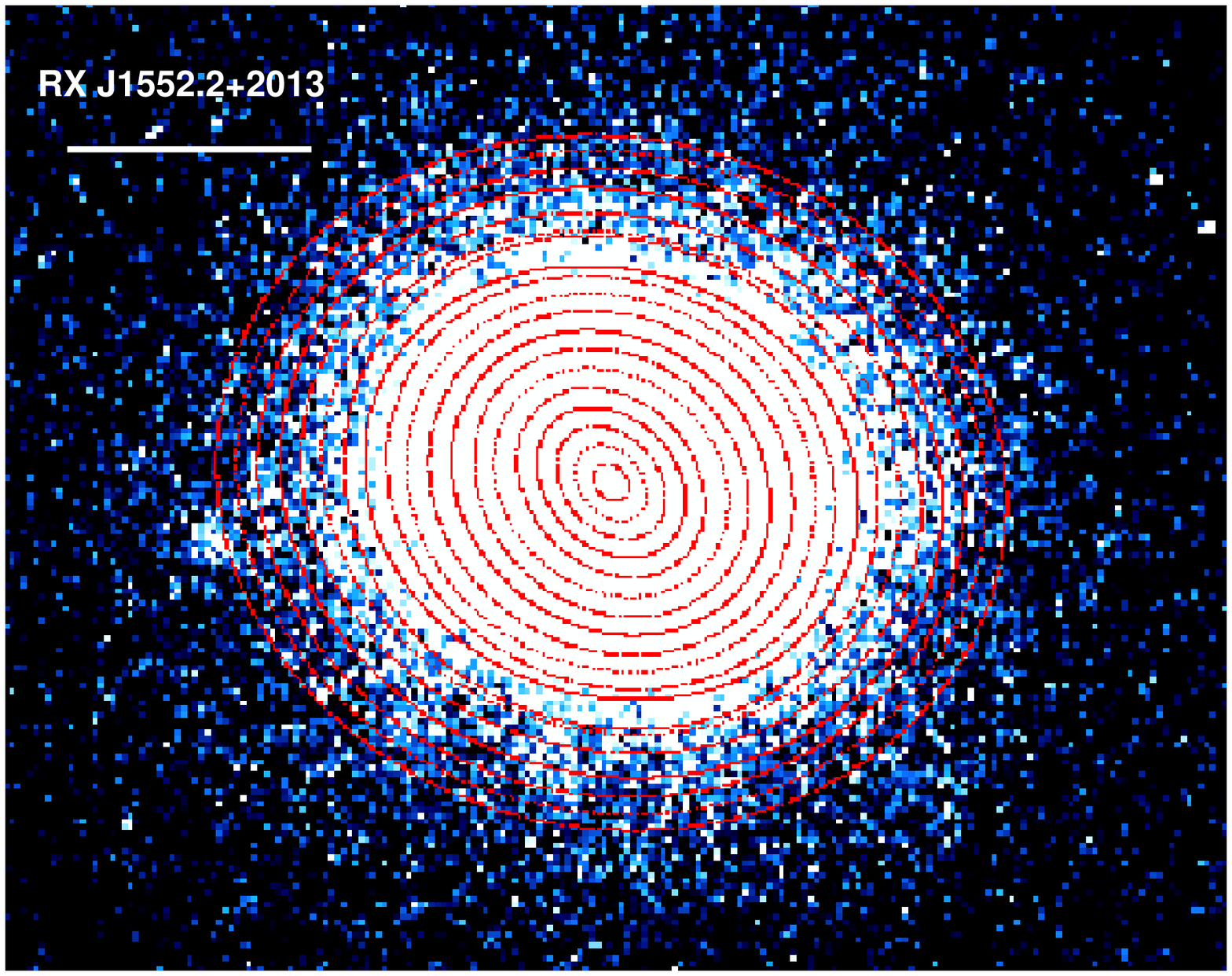,width=1.3in,height=1.2in}
\epsfig{file=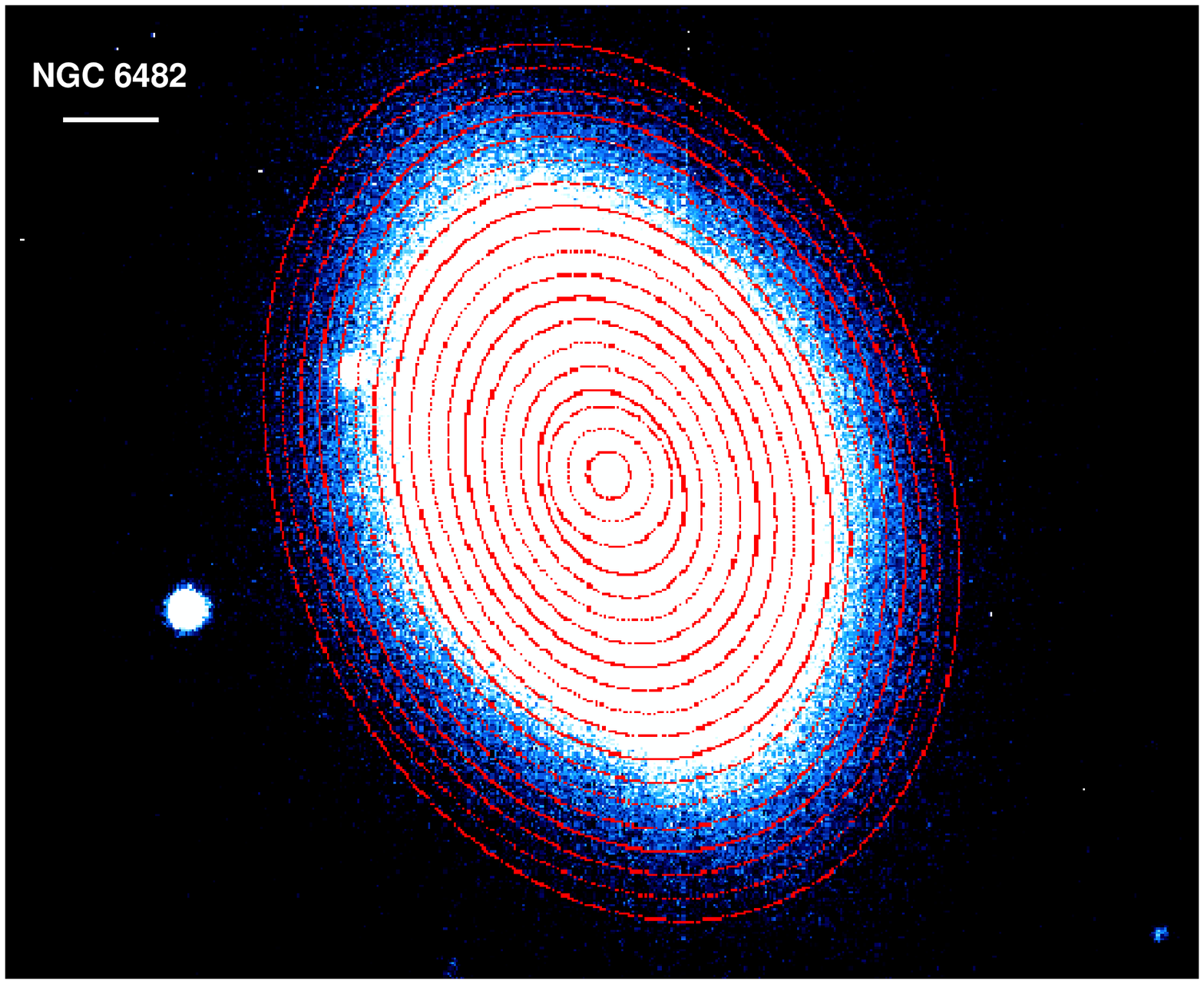,width=1.3in,height=1.1in}
\caption{Ellipse fits to the surface brightness distribution of the 
central galaxy in fossil groups. All the images are from the Ks-band 
observations with the exception of the most distant fossil, RX J1256.0+2556, 
which is observed in R-band. 5 arcsec scale bars are also shown.}
\label{chansoft}
\end{figure}

\section{Fossil galaxy groups}

In the class of galaxy groups known as ``fossil groups'', 
the group is dominated optically by a single luminous elliptical
galaxy at the centre of extended luminous X-ray emission similar to
that seen in bright X-ray groups. The X-ray emission in fossils is
regular and symmetric, indicating the absence of recent group
merging. The dominant giant elliptical galaxy has an
optical luminosity similar to BCGs. A cD galaxy has also been reported
in a fossil group \citep{mendes05}. The observed
properties of fossils, and their absence of $L^\star$ galaxies, suggest
that they must be old galaxy groups. These properties are discussed in 
recent studies \citep{kmpj06,kpj06} where we report a higher dark
matter concentration in fossils, compared to non-fossils groups and
clusters with similar masses, which is consistent with an early
formation epoch.

Observationally a galaxy group is classified as a fossil if \citep{jones03} it
has an X-ray luminosity of $L_{X,bol} \geq 10^{42} h^{-2}_{50}$ erg
s$^{-1}$ spatially extended to few 100 kpc, and the dominant galaxy 
is at least 2 magnitudes brighter (in R-band) than the second ranked galaxy 
within half the projected virial radius of the group. 
The X-ray criterion guarantees
the existence of a group size galaxy halo while the optical criterion
assures that the $M^\star$ galaxies are absent within the given radius
which corresponds to the radius for orbital decay by dynamical friction
\citep{binney87}. No upper limit is 
placed on the X-ray luminosity or temperature, and recently a fossil galaxy 
cluster was found \citep{kmpj06}.

\subsection{The sample}

This study makes use of a flux-limited sample of fossils found in
the catalogue of spatially extended X-ray sources compiled by
Wide Angle ROSAT Pointed Survey (WARPS) project. Details of the 
fossil identification and sample selection is given in \citet{jones03}. 
This is the largest statistical sample 
of fossil groups studied to date. In addition, the nearest known fossil 
group NGC 6482 and the first discovered fossil group, RX J1340.5+4017, are 
included in our sample. The detailed X-ray analysis 
of the sample is the subject of a separate study \citep{kpj06}.

\subsection{Optical and near-IR observations} 

The above sample was observed using the observational facilities of
Issac-Newton Group of Telescopes (ING) and Kitt-peak National
Observatory (KPNO). R-band images were obtained using the INT 2.5m wide
field imager. Unfortunately the conditions were not photometric, and so
further R-band imaging was obtained, in photometric conditions, with the 
8k mosaic camera at the University of Hawaii 2.2-m telescope, and in INT
wide-field camera service time. The resultant photometric accuracy 
for all the systems is $\le$0.05 mag. R-band
observation of NGC 6482 was performed with the KPNO-0.9m in April 2005.
Spectroscopic observations of the sample were also obtained, using slit 
spectroscopy on the KPNO 4m, to examine the optical membership
of these groups. This is discussed in \citet{kpj06}.

$K_s$-band observations of the sample using UIST/UKIRT were performed 
in 2004. The seeing was measured to be $\sim 1.0''$. The
data were reduced using the ORAC data reduction package 
(http://www.oracdr.org/).  
Where multi-snaps were taken, the images were co-added to increase the 
signal-to-noise. Figure \ref{chansoft} shows the $K_s$ images and their 
ellipse fits (section 3.3) for the central fossil galaxies in the sample
with the exception of RX J1256.0+2556 for which only the R-band data was 
available.

\section{Analysis and results}

Elliptical galaxies are usually single-component galaxies with 
radial surface brightness profiles described by a de Vaucouleurs
law ($r^{1/4}$). It has been shown that the Sersic profile ($r^{1/n}$)
gives a better fit, in general, for ellipticals with a wide luminosity range
and in different environments (Trujillo \etal 2001; Khosroshahi \etal 2004). Ellipticals 
are also divided based on their isophotal shapes. For this study we 
concentrate on the radial surface brightness profile,
the ellipticity profile and the fourth order Fourier coefficient
($B_4$), which is an indicator of  boxy and disky isophotes. 
The analysis is performed on both the R-band and $K_s$-band images using 
the well-known IRAF/ellipse task.

\subsection{Radial surface brightness profile}

The Sersic profile, was used to model the R-band surface
brightness distribution of the giant elliptical galaxies, using a two
dimensional bulge/disk decomposition method. The values of $n$ and the
half-light radius, along with other parameters, are given in Table
1. The analysis shows that the fossil central galaxies are best
modelled with $<n>= 4.1 \pm 0.7$. While this agrees in general with 
the surface brightness profiles of the remnants of collisionless
disk mergers, these simulations are not able to produce galaxies
as large as the dominant fossil galaxies \citep{naabtru06}.

Our ground-based observations are inadequate for probing the radial
surface brightness profiles within the central $\sim 1$ kpc, which is 
necessary for a power-law/core classification. 
As a result we limit our investigation to the Sersic fit 
to the galaxy.

\begin{figure}
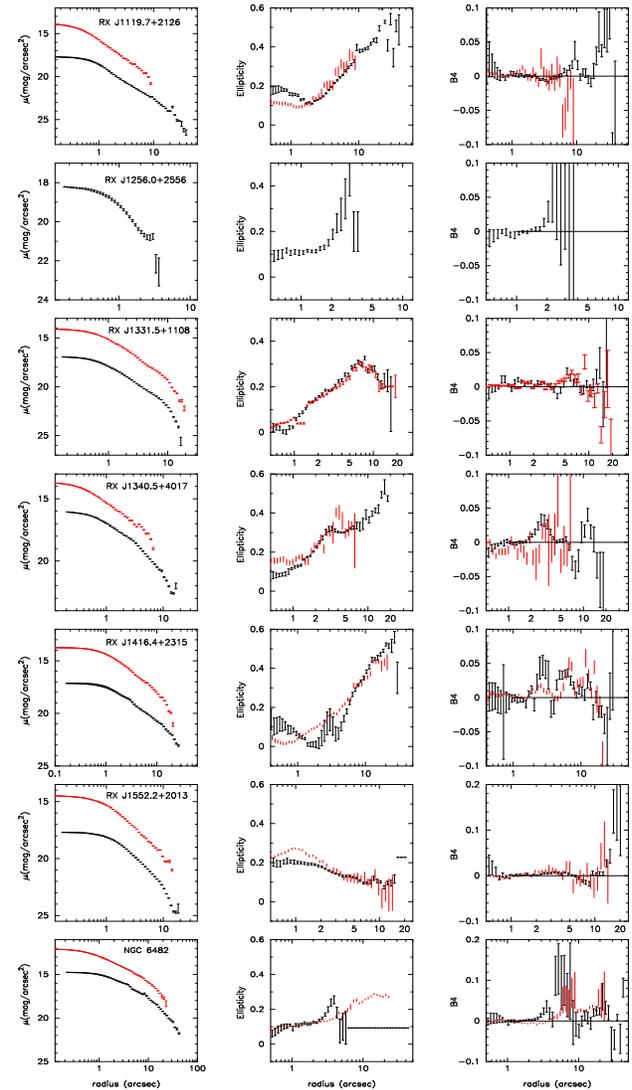

\center
\epsfig{file=isophot-j1119b.ps,width=3.2in,height=0.8in}
\epsfig{file=isophot-j1256b.ps,width=3.2in,height=0.8in}
\epsfig{file=isophot-j1331b.ps,width=3.2in,height=0.8in}
\epsfig{file=isophot-j1340b.ps,width=3.2in,height=0.8in}
\epsfig{file=isophot-j1416b.ps,width=3.2in,height=0.8in}
\epsfig{file=isophot-j1552b.ps,width=3.2in,height=0.8in}
\epsfig{file=isophot-n6482b.ps,width=3.2in,height=0.8in}
\caption{Profiles of radial surface brightness, ellipticity and the 
fourth Fourier coefficient, $B_4$, for  
RX J1119.7+2126, RX J1256.0+2556, RX J1331.5+1108, RX J1340.5+4017, 
RX J1416.4+2315, RX J1552.2+2013 and NGC 6482. The profiles extracted from 
R-band and $K_s$-band data are shown in dark (black) and grey (red), 
respectively.}
\label{isophots}
\end{figure}

\subsection{Ellipticity profile}

The ellipticity profiles presented in Fig \ref{isophots} show a
general pattern in which the ellipticity of the isophotes increases
with the radius, with the exception of RX J1552.2+2013. Galaxies with 
high quality data show
ellipticity increasing to 0.4-0.6 exceeding the ellipticity of the
X-ray halo which is usually less than 0.3 \citep{buote96}. The
central galaxy in fossils are
aligned with the underlying dark matter confirmed both in X-ray and
lensing studies. 
Similar alignment has been noted for the BCGs in many clusters\citep{fuller99}.

\begin{table*}
\begin{center}
\caption{Photometric properties of the sample galaxies.
\label{table1}
}
\begin{tabular}{lllcccccc}
\hline
Group & R.A.    &   Dec. & z &  M$_R$  & $a_4/a$     & $n$ & $r_e$ & kpc/arcsec\\
      & (J2000) &(J2000) &   &  mag    & $\times100$ &     &  arcsec \\
\hline
RX J1119.7+2126 & 11:19:43.6& +21:26:51 & 0.061 &-22.1 & 0.1 & 5.1 & 8.9  & 1.14\\
RX J1256.0+2556 & 12:56:03.4 & +25:56:48 & 0.232 &-24.1 & 0.1 & 3.1 & 7.3  & 3.73\\ 
RX J1331.5+1108 & 13:31:30.2 & +11:08:04 & 0.081 &-22.9 & 0.3 & 4.3 & 7.2  & 1.53\\ 
RX J1340.5+4017$^a$ & 13:40:33.4 & +40:17:48 & 0.171 &-23.0 & irr & 4.2 & 7.6  & 2.92\\
RX J1416.4+2315 & 14:16:26.9 & +23:15:32 & 0.137 &-24.3 & 0.7 & 3.6 & 12.1 &2.44\\
RX J1552.2+2013 & 15:52:12.5 & +20:13:32 & 0.135 &-24.0 & 0.5 & 4.6 & 15.8 &2.40\\
NGC 6482$^b$ & 15:52:12.5 & +20:13:32    & 0.013 &-22.9 & 1.3 & 3.8 & 16.0 &0.26\\

\hline
\end{tabular}
\end{center}
$^a$ This system is the first confirmed fossil group and not part of the 
flux-limited sample of fossils. $^b$ This group is  known to be the nearest fossil system 
\citep{kjp04} and is not part of the flux-limited sample.
\end{table*}

\subsection{Isophotal analysis}

Fig \ref{isophots} shows the results of the ellipse fits.
In order to quantify the shape
of the isophotes and to be able to make a direct comparison with
similar analyses in the literature, we calculate $a_4/a$, which is based on
the measured fourth Fourier coefficient, $B_4$ \citep{jorgensen99}. 
Similarly to \citet{bender89}, $a_4/a=\frac{\sqrt{1-\epsilon}B_4} 
{adI/da}$ is quantified at its peak value. In the absence of a peak
the $a_4/a$ is quantified at $r_e$. 
Here $\epsilon$, $a$ and $I$ are the ellipticity, semi-major axis length 
and the surface brightness of the isophotes, respectively. 
As seen in Fig.\ref{isophots} none of the galaxies have predominantly  
boxy isophotes, except in the outskirts were the statistics are 
very poor and the values of $B_4$ are not well constrained.

\begin{figure}
\center
\epsfig{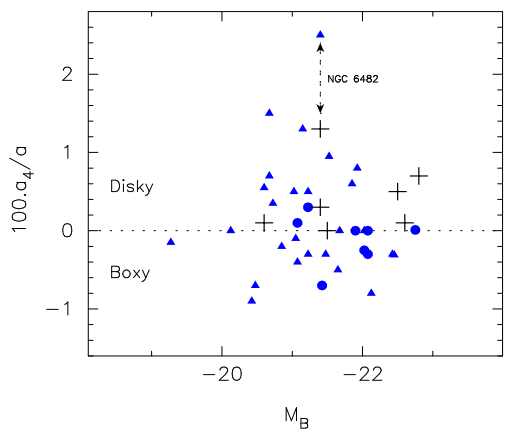}
\caption{The variation in isophotal shapes of early-type 
brightest group or cluster galaxies, and 
those in fossil groups (crosses), with the optical luminosity of the brightest
galaxy. The comparison sample is a combination of early-type 
BGGs (triangles) and BCGs (circles) from \citet{ellis06}, for 
which $a_4/a$ values were available.}
\label{a4}
\end{figure}

We compare the values of $a_4/a$ with those of the brightest galaxies in 
groups and clusters, BGGs and BCGs (Fig \ref{a4}). It is clear from this 
plot that none of the fossil's dominant galaxies have prominent boxy 
isophotes. Indeed some, including the nearest fossil, NGC 6482 \citep{kjp04}, 
and the fossil cluster, RX J1416.4+2315 \citep{kmpj06}, are highly disky 
isophote galaxies. An earlier study \citep{faber97} gives an even higher 
value for the diskyness of NGC 6482.  The comparison sample is a 
combination of early-type BGGs and BCGs from \citet{ellis06} for 
which $a_4/a$ values were available from earlier studies and therefore 
it is not a complete sample.

\section{Discussion and conclusions}

This analysis shows that, despite apparent similarities, the dominant
giant elliptical galaxy in fossil groups are different in their
isophotal shapes, compared to the brightest central
galaxies in non-fossil systems, especially in rich clusters. Luminous 
elliptical galaxies in non-fossil groups and clusters do not 
present disky isophotes. Less luminous BGGs
show disky and boxy isophotes in similar 
proportions. However, the observed frequency of boxy-isophote dominant fossil 
galaxies is apparently zero. 

If the central galaxies of fossil groups have indeed been formed from the
merger of all major galaxies within the inner regions of the group, as
we suppose, then some of these mergers would have been gas-rich, as
group spirals are incorporated into the central merger-remnant.
The disky character of central fossil galaxies is
then consistent with the findings of numerical simulations, that
disky isophotes result from gas rich mergers.
\citet{khochfar05} highlight the importance of the role of gas 
in galaxy mergers and show that the isophotal shapes of merger
remnants are sensitive to the morphology of their progenitors and
to subsequent gas infall. 
In contrast, boxy isophote ellipticals 
are formed by equal-mass mergers of bulge dominated galaxies. Such 
mergers are likely to occur at the core of clusters where most of the 
galaxies are gas poor.

Almost 90\% of BCGs studied by \citet{laine03} are core galaxies, ie.
with flatter slope near the nucleus in their radial surface brightness 
profiles. The 
study by \citet{faber97} shows that a large fraction ($\sim$70\%) of core 
galaxies have boxy isophotes, with as low as $\sim$10\% with disky isophotes, 
implying a strong association of boxy isophotes with core galaxies. 
Taking into account the conversion of cuspy cores into flat
low-density cores by black hole merging, \citet{khochfar05} find that
disky ellipticals should contain central density cusps whereas boxy
ellipticals should in general be characterised by flat cores. Only
rare low-luminosity boxy ellipticals, resulting from equal-mass
mergers of disk galaxies, could have power-law cores. 
Space resolution data is needed to study the core of the galaxies to
verify the core and power- law property of dominant fossil galaxies.
If future observations show that fossil group dominant galaxies are
power-law galaxies, as expected from their isophotes in light of the
above argument, then they will be the first of such entities to grow
to the size of BCGs.

\begin{figure}
\center
\epsfig{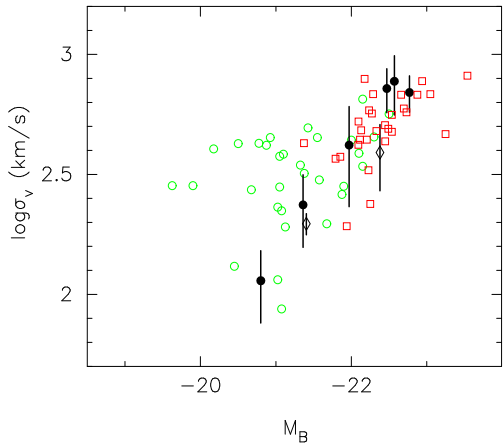}
\caption{A comparison between the correlation of group/cluster
velocity dispersion and absolute B-magnitude of the dominant
galaxy. The data points from Table 1 (except RX J1119.7+2126) are
shown with filled circles. Diamonds represent two of the fossil candidate 
OLEGs, out of four, in \citet{yoshioka04}. 
The correlation is much tighter in fossils than in non-fossil
groups (circles) and clusters (squares). The comparison non-fossil
groups are GEMS X-ray selected groups with early-type BGG and with
group scale X-ray emission (G-sample in \citet{osmond04}). The cluster
sample is selected from \citet{girardi02} for which the luminosity 
of the central galaxy was available in \citet{mohr04}.}
\label{sigma}
\end{figure}

The link between fossil galaxy groups and the BCGs is further
motivated by observational space density estimates of fossils which is
found to be 8\% to 20\% of X-ray luminous systems \citep{jones03} and 
as large as the space density of poor and rich galaxy clusters combined. 
This means that there are enough fossils to provide BCGs to clusters. 
A recent theoretical study \citep{milos06} predicts 5\%-40\% of galaxy 
groups and 1\%-3\% of galaxy clusters to be fossils. In the context of BCG 
formation, fossils appear to be suitable environments for the formation of 
luminous giant ellipticals around which clusters 
can form. This is a different formation mode to the one in 
which the BCG forms via mergers of brightest group galaxies during the 
cluster collapse. 
Numerical simulations suggest that the isophotes of the BCGs formed in the
latter case will be predominantly boxy, characteristic of gas free
(``dry'') mergers. 

The BGGs and BCGs with non-boxy isophotes in Fig.\ref{a4} could easily 
originate as fossil group central galaxies which have been incorporated
into larger structures. The difference in photometric structure seen
clearly in many BCGs and BGGS, especially the most luminous ones,
does not rule out the possibility that they originated in fossil groups.
This could still be the case provided that they have undergone later
gas-free mergers within the cluster environment -- for example, with
the BGGs of infalling galaxy groups. 
About 40\% of BCGs contain at least one secondary nucleus\citep{laine03},
which strongly suggests the action of late mergers within the 
cluster environment. If fossils are indeed old and undisturbed systems 
then they should be found to have a very {\it low} incidence of 
multiple nuclei. 
 
In support of the above argument we show (Fig \ref{sigma}) a tight
correlation between the luminosity of the central galaxy in fossils
and the underlying gravitational mass probed by the group velocity
dispersion.  This shows that fossils form a homogeneous population
in which the luminosity of the central galaxy is strongly tied to the
property of its parent group. This property of the fossils is
consistent with their early formation epoch and absence of recent
merger.  In contrast the large scatter in the distribution of the
non-fossil galaxy groups and clusters, on the same plane, is
understood to be merger driven. Absence of a recent major merger helps 
the dominant fossil galaxy preserve its original structure.  

We conclude that there is high chance that disky 
BCGs are formed in fossil groups. Boxy BCGs, could still result from
progenitor fossils, but would need to have
undergone a dry merger within the cluster, probably
as a result of merger with BGGs of infalling groups.

We would like to thank Doug Burke, for his involvement in the 
UH 88'' observations, and the INT for the service observations,  
Ewan O'Sullivan for his involvement in the KPNO observations
of  NGC 6482, Graham P. Smith for his valuable input 
which helped to improve the discussion, and Arif Babul for his 
useful comments.

\bsp

\label{lastpage}

\begin{thebibliography}{99}
\bibitem[\protect\citeauthoryear{Barnes}{1989}]{barnes89}
Barnes J. E., 1989, Nat, 338, 123
\bibitem[\protect\citeauthoryear{Bender}{1988}]{bender88}
Bender, R. 1988, A\&A, 193, 7
\bibitem[\protect\citeauthoryear{Bender}{1989}]{bender89}
Bender R., Surma P., Doebereiner S., Moellenhoff C., Madejsky R., 
1989, A\&A, 217, 35
\bibitem[\protect\citeauthoryear{(Bender, Burstein \& Faber}{1992}]{bender92}
Bender R., Burstein D., Faber S. M., 1992, ApJ, 399, 462
\bibitem[\protect\citeauthoryear{Binney \& Tremaine}{1987}]{binney87}
Binney K., Tremaine S., 1987, Galactic Dynamics. Princeton Univ. Press, 
Princeton, NJ
\bibitem[\protect\citeauthoryear{Buote \& Canizares}{1996}]{buote96}
Buote D. A., Canizares C. R., 1996, ApJ, 457, 565
\bibitem[\protect\citeauthoryear{(Dressler}{1980}]{dressler80}
Dressler A., 1980, ApJ, 236, 351
\bibitem[\protect\citeauthoryear{Dubinski}{1998}]{dubinski98}
Dubinski, J. 1998, ApJ, 502, 141)
\bibitem[\protect\citeauthoryear{Ellis \&O'Sullivan}{2006}]{ellis06}
Ellis S. C., O'Sullivan E., 2006, MNRAS, 367, 627
\bibitem[\protect\citeauthoryear{Emsellem \etal}{2004}]{emsellem04}
Emsellem E., Cappellari M., Peletier R. F., McDermid R. M. 2004, MNRAS, 352, 721
\bibitem[\protect\citeauthoryear{Faber \etal}{1997}]{faber97}
Faber S. M., Tremaine S., Ajhar E. A., Byun Y. \etal, 1997, AJ, 114, 1771
\bibitem[\protect\citeauthoryear{Fuller, West \& Bridges}{1999}]{fuller99}
Fuller T. M., West J. W., Bridges T. J., 1999, 519, 22
\bibitem[\protect\citeauthoryear{Girardi \etal}{2002}]{girardi02}
Girardi M., Manzato P., Mezzetti M., Giuricin G., Limboz F., 2002, ApJ, 569, 720
\bibitem[\protect\citeauthoryear{Trujillo \etal}{2001}]{trujillo01}
Trujillo I., Graham A. W. and Caon N. 2001, MNRAS, 326, 869
\bibitem[\protect\citeauthoryear{Hausman \& Ostriker}{1978}]{hausman78}
Hausman, M. A., \& Ostriker, J. P. 1978, ApJ, 224, 320
\bibitem[\protect\citeauthoryear{Jorgensen \etal}{1999}]{jorgensen99}
Jorgensen I., Franx, M., Hjorth, J., van Dokkum P. G., 1999, MNRAS, 308, 833 
\bibitem[\protect\citeauthoryear{Jones \& Forman}{1984}]{jf84}
Jones, C., \& Forman, W. 1984, ApJ, 276, 38
\bibitem[\protect\citeauthoryear{Jones \etal}{2003}]{jones03}
Jones L. R., Ponman T. J., Horton A., Babul A., Ebeling H., Burke D. J.,
2003, MNRAS, 343, 627
\bibitem[\protect\citeauthoryear{Jones \etal}{2000}]{jones00}
Jones L. R., Ponman T. J., Forbes D.A., 2000, MNRAS, 312, 139
\bibitem[\protect\citeauthoryear{Khochfar \& Burkert}{2005}]{khochfar05}
Khochfar S., Burkert A., 2005, MNRAS, 359, 1379
\bibitem[\protect\citeauthoryear{Khosroshahi \etal}{2004}]{habib04}
Khosroshahi H. G., Raychaudhury S., Ponman T. J., Miles T. A. Forbes D.,  
2004, MNRAS, 349, 524
\bibitem[\protect\citeauthoryear{Khosroshahi, Jones \& Ponman}{2004}]{kjp04}
Khosroshahi H. G., Jones L. R. and Ponman T. J., 2004, MNRAS, 349, 1240
\bibitem[\protect\citeauthoryear{Khosroshahi \etal}{2006}]{kmpj06}
Khosroshahi H. G., Maughan B., Ponman T. J., and Jones L. R., 2006, 
MNRAS, to be published (astro-ph/0603606)
\bibitem[\protect\citeauthoryear{Khosroshahi, Ponman \& Jones}{2006}]{kpj06}
Khosroshahi H. G., Ponman T. J., and Jones L. R., 2006, MNRAS, submitted
\bibitem[\protect\citeauthoryear{Khosroshahi \etal}{2000}]{kwkm00}
Khosroshahi H. G., Wadadekar Y., Kembhavi A., Mobasher B., 2000, ApJ, 531, L103
\bibitem[\protect\citeauthoryear{Laine \etal}{2003}]{laine03}
Laine S., \etal, 2003, AJ, 125, 478
\bibitem[\protect\citeauthoryear{Mendes de Oliveira, Cypriano \& Sodre Jr.}
{2005}]{mendes05}
Mendes de Oliveira C., Cypriano E. S., Sodre Jr. L., 2005, astro-ph/0509884
\bibitem[\protect\citeauthoryear{Milosavljevic \etal}{2006}]{milos06}
Milosavljevic M., Miller C. J., Furlanetto S. R., Cooray A., 2006, ApJ, 637, L9
\bibitem[\protect\citeauthoryear{Lin \& Mohr}{2004}]{mohr04}
Lin Y.-T. \& Mohr J. J., 2004, ApJ, 617, 879
\bibitem[\protect\citeauthoryear{Naab \& Burkert}{2003}]{naab03}
Naab T., Burkert A., 2003, ApJ, 597, 893
\bibitem[\protect\citeauthoryear{Naab, Khochfar \& Burkert}{2006}]{naab06}
Naab T., Khochfar S., Burkert A., 2006, ApJ, 636, L81
\bibitem[\protect\citeauthoryear{Naab \& Burkert}{2003}]{naabtru06}
Naab T., Trujillo I., 2006, MNRAS, 369, 625
\bibitem[\protect\citeauthoryear{Osmond \& Ponman}{2004}]{osmond04}
Osmond J.P.F., Ponman T.J., 2004, MNRAS, 350, 1511	
\bibitem[\protect\citeauthoryear{Ponman \etal}{1994}]{ponman94}
Ponman T. J., Allan D. J., Jones L. R., Merrifield M., MacHardy I. M., 1994,
Nature, 369, 462
\bibitem[\protect\citeauthoryear{Rest \etal}{2001}]{rest01}
Rest A., \etal 2001, AJ, 121, 2431
\bibitem[\protect\citeauthoryear{Scorza \& Bender}{1995}]{scorza95}
Scorza C., Bender R., 1995 A\&A, 293, 20	
\bibitem[\protect\citeauthoryear{Searle, Sargent \& Bagnuolo}{1973}]{searle73}
Searle L., Sargent W. L. W., and Bagnuolo W. G., 1973, ApJ, 179, 427
\bibitem[\protect\citeauthoryear{Smith \etal}{2005}]{smith05}
Smith G. P. Kneib J., Smail I., Mazzotta P., Ebeling H., Czoske, O. , 2005,
MNRAS, 359, 417
\bibitem[\protect\citeauthoryear{Toomre \& Toomre}{1972}]{toomre72}
Toomre A., and Toomre J., 1972, ApJ, 178, 623
\bibitem[\protect\citeauthoryear{Sun \etal}{2003}]{sun03}
Sun M., Forman W., Vikhlinin A., Hornstrup A., Jones C., Murray S. S., 2003, ApJ, 598, 250
\bibitem[\protect\citeauthoryear{Yoshioka \etal}{2004}]{yoshioka04}
Yoshioka T., Furuzawa A., Takahashi S., Tawara Y., Sato S., Yamashita K.,
Kumai Y., 2004, Adv. in Space Res, 34, 2525

\end{thebibliography}
\end{document}